\documentclass[reprint,flushbottom,showpacs]{revtex4-1}

\usepackage{psfrag} % Usage: %\psfrag{tag}[position][psposition][scale][rotation]{LaTeX construction}
\usepackage{graphicx}
\usepackage[usenames]{color}	% use color by calling color_names

\usepackage{amsmath}
\usepackage{amssymb}
\usepackage{amsthm}
\usepackage{bm}
\usepackage{color}

\newcommand{\bea}{\begin{eqnarray}}
\newcommand{\ea}{\end{eqnarray}}
\newcommand{\eea}{\end{eqnarray}}
\newcommand{\ord}{{\cal O}}
\newcommand{\sumint}[1]

\usepackage{graphicx}			% graph
\usepackage{amssymb,amsmath}	% from aps
\usepackage{hyperref}			% refference and URL
\newcommand{\be}{\begin{equation}}
\newcommand{\ee}{\end{equation}}
\newcommand{\bdm}{\begin{displaymath}}
\newcommand{\edm}{\end{displaymath}}
\newcommand{\bpm}{\begin{pmatrix}}
\newcommand{\epm}{\end{pmatrix}}
\newcommand{\bc}{\begin{center}}
\newcommand{\ec}{\end{center}}
\begin{document}

%\title{Truncated many-body dynamics of interacting bosons with monitored error}
% and condensate fragmentation} % of Bose-Einstein condensates}
\title{Truncated many-body dynamics of interacting bosons: \\ A variational principle with error monitoring}
\author{Kang-Soo Lee and Uwe R. Fischer}
\affiliation{Seoul National University, Department of Physics and Astronomy\\ Center for Theoretical Physics, 151-747 Seoul, Korea }
\date{\today}

\begin{abstract}
We develop a method to describe the temporal evolution of an interacting system of bosons,
for which the field operator expansion is truncated after a finite number $M$ of modes, in a rigorously controlled manner. 
Using McLachlan's principle of least error, we find a self-consistent set of equations for the many-body state.
As a particular benefit, and in distinction to previously proposed approaches,  the presently introduced method facilitates the dynamical increase of the number of orbitals during the temporal evolution, due to the fact that we can rigorously monitor the error made by increasing the truncation dimension $M$.
The additional orbitals, determined by the condition of least error of the truncated evolution relative to the exact one, 
are obtained from an initial trial state by steepest {\em constrained} descent.
\end{abstract}

\maketitle

%\tableofcontents
%\newpage

%%%%%%%%%%%%%%%%%%%%%%%%%%%%
\section{Introduction}

Since the experimental realization of Bose-Einstein 
condensates \cite{Anderson,Davis,Bradley}, a large variety of experiments with bosonic isotopes, either atoms or molecules, have opened up a fascinating mesoscopic and macroscopic quantum world \cite{Stamper}. 
%To understand these experimental results theoretically and to render results systematically computable, the %second quantization formalism is commonly used. 
After an initial period, concentrating on the effective single-particle physics of these ultracold dilute gases
\cite{Dalfovo}, more recently their many-body physics came into focus, revealing rich and hitherto unexpected possibilities to test fundamental correlation properties at the microscopic level \cite{Bloch}. There is a plethora of phenomena to be explored, for example by going beyond the conventional contact interaction pseudopotential to long-range, in particular dipolar, interactions \cite{Baranov}, when placing the gas in an optical lattice, mimicking certain aspects of the behavior of electrons in solids \cite{Lewenstein}, or when the intricate correlations of interacting many-body systems far from equilibrium are studied \cite{Dziarmaga}.
The fundamental quest into quantum many-body physics has also been, inter alia, stimulated by the promise offered through {\em quantum simulation} \cite{Nori,Georgescu}, i.e., to employ the highly 
controllable ultracold dilute quantum gases to study other less controllable or even inaccessible quantum systems. 
The latter frequently occur in solid state physics, which is plagued by various practical problems, e.g.,  
sample preparation within exact specifications.

The extension to a true many-body physics, that is, incorporating quantum correlations beyond mean-field essentially to any order, requires, however, vast computational resources when both the number of particles and the interactions increase. 
Therefore, a simplification of the problem by {\em truncating} the field operator expansion to a finite number of modes (or, as an equivalent term,  single-particle orbitals) has been commonly utilized to obtain results relevant to the prediction of experiments in trapped bosonic quantum gases. 
The most extreme truncation, the semiclassical form of mean-field theory, retaining just one orbital, gives the well-known Gross-Pitaevski\v\i\/ equation. 
Without the aid of contemporary computers, it seemed to be inevitable until most recently, particularly in out-of-equilibrium situations far away from the ground state, to reduce the complexity of the problem at hand as much as possible, and hence to use the Gross-Pitaevski\v\i\/ equation approach. With the increased interest in many-body physics, however, there arose the necessity to go beyond the all-too-simplified mean field approach of the  Gross-Pitaevski\v\i\/ equation. The accuracy of predictions on many-body correlations and the corresponding response functions will %obviously 
increase with a less severe degree of truncation. %, though the solutions will not be exact still.
% Necessity of truncation of field operator expansion to render problems computable

To derive the equations of many-body evolution, various variational approaches can be employed. 
Historically the first was the variational ansatz of Dirac and Frenkel \cite{Dirac, Frenkel}, followed by  McLachlan's variational principle \cite{McLachlan} and the time-dependent variational principle (TDVP), which is a principle of stationary action \cite{Broe, MCTDHB}. Therefore, there are various, not necessarily equivalent, choices of variational principle for finding the equation of motion of the truncated many-body evolution. The Dirac-Frenkel principle imposes $\langle\delta\Phi| \hat{H}-i\partial_t |\Phi\rangle = 0$ 
($\hbar\equiv 1$), where $\langle\delta \Phi|$ denotes any possible variations of the many-body state $\langle\Phi|$ with respect to a given set of variational parameters, whereas McLachlan's principle requires that the error of many-body evolution must be minimized. 
%\col{Sometimes you use $\Psi$, sometimes $\Phi$ for the many-body state; is there supposed to be a difference, for example one is truncated the other exact? If not, we should use one symbol, e.g. $\Phi$ and its variation.}  
On the other hand, the TDVP, as stated, requires stationarity of a given action. The three principles thus support quite different doctrines. 

Applying either the TDVP or the Dirac-Frenkel principle, in \cite{MCTDHB,MCTDHB1, MCTDHB2, MCTDHB3} a method  coined MCTDHB (Multi-Con\-figura\-tional Time-Dependent Hartree method for Bosons) has been proposed. 
%Inter alia, the latter // Not the latter one. Both principles gave the same result in their paper.
This approach has, for example, provided tools for the description of macroscopic condensate fragmentation 
of bosonic many-body states \cite{Penrose}, which is of nontrivial relevance in particular for the description
of fragmented condensates in single traps \cite{Bader,Fischer,Kang}. 
We will describe below in detail that, besides its many beneficial properties and numerical successes, the MCTDHB method remains incomplete in certain situations. Specifically, when the single-particle density matrix (SPDM) becomes singular, i.e., noninvertible, the method fails to provide a self-consistent solution, and has to be
repaired by hand. As a consequence, MCTDHB does not provide a recipe to propagate, for example, a single condensate into a fragmented condensate many-body state. Although MCTDHB provides an important tool to describe the many-body physics of interacting bosons, the method therefore lacks the possibility to directly connect the phenomena of condensation and fragmentation. 

%will show: ambiguity of [previously employed Dirac-Frenkel variational principle for many-body evolution
%MacLachlan principle offers possibility to constrain error accumulated during evolution
%demonstrate that during time evolution SPDM(Single Particle Density Matrix) potentially becomes singular (noninvertible), 

Here, critically examining the Dirac-Frenkel principle and the TDVP, and then adopting alternatively McLachlan's principle for truncated many-body evolution, we improve on the previously proposed multi-configurational Hartree methods, in that we propose a procedure for the solution of the singularity problem of a noninvertible SPDM. We will also validate the resulting equations of MCTDHB in a different manner, however, additionally offering a straightforward handling of the exceptional evolution points related to the singularity of the SPDM. 

%invalidating Dirac-Frenkel principle 
%mention cases of noninvertibility 

%our scheme offers possibility to increase/decrease number of orbitals in truncated field
%operator expansion in a controlled manner, i.e. with a monitored error

%%%%%%%%%%%%%%%%%
\section{Variational Principles}

Let us now discuss the possible variational principles in more detail.
We are aiming at finding an approximate solution of the many-body Schr\"odinger equation when the state $|\Phi\rangle$ is restricted (or truncated). McLachlan's principle \cite{McLachlan}, which was presented in 1963 as a new version of Frenkel's principle, requires the minimization of the error or remainder of this approximate solution from the exact evolution. 
The time evolution of any state is dictated by Schr\"odinger's equation, 
	$i\partial_t |\Phi\rangle = \hat{H} |\Phi\rangle$. 
	In other words, the evolution of state is determined by the Hamiltonian at any moment. However, to make the state $|\Phi\rangle$ manipulable, we are generally forced to restrict or confine the state $|\Phi\rangle$ into some simple and computationally feasible forms. With the state $|\Phi\rangle$ in restricted form, $[i\partial_t - \hat{H}] |\Phi\rangle$ cannot be exactly zero in general. Therefore, McLachlan's principle aims at finding the approximate solution which minimizes the positive semidefinite error measure 
	$\langle\Phi| [i\partial_t - \hat{H}]^{\dag}[i\partial_t - \hat{H}] |\Phi\rangle$. The details of the corresponding procedure will be rephrased in section \ref{subsec: Adapting the number of orbitals}, after introducing a concrete way to restrict the many-body state in a computationally feasible form.

Hence it is guaranteed that the equation of motion obtained from McLachlan's principle follows the exact evolution most similarly under given constraints. The most appealing feature of McLachlan's principle is thus that it is  a quite  intuitive principle. 
Since it offers the possibility to evaluate the error directly, we can intermediately, monitoring the error, increase the number of orbitals in the truncated field operator expansion, i.e. truncate the state less, to assure accuracy of the result. Alternatively, to save computational costs and time, the number of orbitals can also be decreased intermediately, i.e. by truncating the state more, in particular in cases where decreasing the number of orbitals does not affect the accuracy of the result significantly. 
Our scheme, described in detail below, in which McLachlan's principle is applied, therefore offers the opportunity to dynamically adjust the truncation of field operators or the state itself properly during computational time evolution, since we can monitor the error. As a particular benefit, a continously applied convergence test, mandatory for MCTDHB, is unnecessary, as we can directly obtain an error which indicates automatically how well our approach describes the interacting system of bosons. 
In addition, the TDVP which was carried out in MCTDHB \cite{MCTDHB} requires stationarity of action, $\delta S=0$. This does not necessarily mean an extremum (minimum or maximum) of the action. Though stationary points include local extrema, the principle practically imposes only stationarity of the action: The equation of motion comes from a stationary point of the action, which is not even necessarily a local minimum or maximum. 

%$\delta S=0$ does not necessarily give extremum

In many-body quantum mechanics, the action is written in terms of an expectation value of an operator-valued functional: 
\begin{widetext}
\be \label{Action}
\begin{split}
	S =
	&\int \mathrm{d} t \int \mathrm{d} \vec{x} ~
	\big\langle \Phi \big|
	\bigg[
	\frac{1}{2m}
	\sum_{i=1}^{3}
	\big( \partial_i \hat{\Psi}^{\dag} (\vec{x}) \big)
	\big( \partial_i \hat{\Psi} (\vec{x}) \big)
	+
	V_{\rm trap} (\vec{x},t)
	\hat{\Psi}^{\dag} (\vec{x}) 
	\hat{\Psi} (\vec{x})
	\bigg]
	\big| \Phi \big\rangle \\ %%%%%
	&+
	\frac{1}{2}
	\int \mathrm{d} t 
	\iint \mathrm{d} \vec{x}_{\alpha} 
	\mathrm{d} \vec{x}_{\beta} ~
	\big\langle \Phi \big|
	V (\vec{x}_{\alpha}, \vec{x}_{\beta})
	\hat{\Psi}^{\dag} (\vec{x}_{\alpha}) 
	\hat{\Psi}^{\dag} (\vec{x}_{\beta})
	\hat{\Psi} (\vec{x}_{\beta}) 
	\hat{\Psi} (\vec{x}_{\alpha})
	\big| \Phi \big\rangle
	-
	\int \mathrm{d} t ~
	\big\langle \Phi \big|
	\bigg[
	\frac{\big[ i \partial_t \big]
	+ \big[ i \partial_t \big]^{\dag}}
	{2}
	\bigg]
	\big| \Phi \big\rangle ,
\end{split}
\ee
\end{widetext}
which is in quantum mechanical correspondence to the classical Lagrangian action. 
Here, $V_{\rm trap}(\vec{x},t)$ is the (in general time-dependent) scalar trap potential confining the atoms, 
$V (\vec{x}_{\alpha}, \vec{x}_{\beta})$ is the two-body interaction potential, and $m$ the mass of bosons. 
This is a real-valued functional of the many-body state $| \Phi \rangle$ \cite{Kull}. Sometimes the action is simply expressed as 
	$S = \int \mathrm{d} t ~
	\langle \Phi | 
	\hat{H} - i \partial_t 
	| \Phi \rangle$. %But this is not a proper expression. As the Lagrangian $L(q,\dot{q};t)$ in classical mechanics is variable-sensitive, the proper form of the action in many-body quantum mechanics is important problem. 
	Variationally changing the state $| \Phi \rangle$ and the temporal change of the state $\partial_t | \Phi \rangle$, we find the evolution of the state around the stationary action point. 

However, when the form of the state is restricted or truncated for computational reasons in the sense that the state resides in a sub-Hilbert space, it is questionable whether the equation of motion obtained in the sub-Hilbert space leads to an evolution most similarly to the exact one obtained with the unlimited full Hilbert space. Even though the equation of motion obtained variationally with a non-truncated state can give the correct many-body Schr\"odinger equation, we cannot rely on the correctness of the equation when the state is truncated. For example, the path of stationary action with limitations imposed on the path can in principle deviate far from the one obtained without any constraints on the path.

As a simple specific example, when we restrict the state to have only an overall phase change, i.e. $| \Phi (t) \rangle = e^{-i \Omega t} | \Phi \rangle$, the action becomes $S = \int \mathrm{d} t \big( \langle \hat{H} \rangle - \Omega \big)$. Depending on the value of $\Omega$, the action can be positive or negative. Actually there is no upper bound and lower bound on this action. For some types of constrained states as above, there can be no stationary point of the action at all. So when applying the TDVP, at least the convergence upon increasing the number of orbitals, i.e. loosening the constraints, must be tested for every specific problem, to ensure the validity of the results. This is because, in contrast with McLachlan's principle, there is no direct error indicator in the TDVP which controls the accuracy of the approximation.

The earliest variational principle for the approximate solution of many-body dynamics is Dirac-Frenkel's principle \cite{Dirac, Frenkel}, which requires 
\be
	\langle\delta\Phi| 
	\hat{H}-i\partial_t 
	|\Phi\rangle = 0 ,
\ee
where $\delta \Phi$ denotes possible variations of the many-body state $\Phi$ with respect to the variational parameters. The equation is quite similar to the TDVP when the action is given by 
	$S
	= \int \mathrm{d}t
	\langle\Phi |
	\hat{H} - i \partial_t
	|\Phi\rangle$. 	
The difference and (possible) equivalences between Dirac-Frenkel's, McLachlan's and other variational principles
have been extensively discussed in the past \cite{Broe, Raab}. 
In \cite{Broe}, it is concluded that if the relevant manifold, i.\,e. the sub-Hilbert space, can be parametrized by pairs of complementary parameters, the above mentioned principles are equivalent. In \cite{Raab}, it is insisted that both Dirac-Frenkel's and McLachlan's variational principles are equivalent if both $\delta \Phi$ and $\delta \Phi^*$ are possible independent variations. But ``equivalence'' here merely indicates the same resulting equation under given conditions, not the equivalence of the principles themselves. 
In addition, we note that the simple-minded point of view that 
$\delta \Phi$ and $\delta \Phi^*$ are possible independent variations can easily lead to incorrect conclusions,  
 since $\delta \Phi^*$ is simply the complex conjugate of $\delta \Phi$ \cite{footnote0}.
% and therefore $\delta \Phi^*$ is just dependent on $\delta \Phi$. 
Furthermore, as explained in detail later, principles which result in the problem of a noninvertible SPDM, which was mentioned already in the above, lack some information in comparison to McLachlan's principle, which resolves this problem.

%These three principles seem to be different in many senses. First of all, $\delta \Phi$ and $\delta \Phi^*$ are quite dependent as they are simply complex conjugate of each other.

% Broe, CPL 149, 547 (1988): We show that if the manifold can be parameterized by pairs of “complementary” parameters, the abovementioned principles are equivalent. The condition of complementarity, which is a sufficient one, is demonstrated to be satisfied in a number of important applications. However, a case of non-equivalence in the recent literature warns against liberal assumptions about the equivalence of the time-dependent variational principles.

% Without any confinement in a state, both give the correct time-dependent Schr\"odinger equation of many-body state. But we should check carefully whether, whatever restrictions and conditions come to a state, both give the same equations and equivalent results. Here we are going to claim that they are not equivalent when the reduced SPDM $\langle \hat{a}_{ij}^{\dag} \rangle$ becomes noninvertible. \\

In summary, comparing the three variational principles, McLachlan's principle appears to be most suitable for finding a truncated many-body dynamics which approximates the real dynamics of interacting bosons. %and is thus automatically rigorously controlled. 
Adopting McLachlan's principle, in the following sections, we demonstrate that the variationally optimal truncation of the many-body dynamics of interacting bosons can be adaptively controlled with a monitored error.

%Dirac-Frenkel's principle seems to have little connection to other physical principles and phenomena such as Hamilton's principle in classical mechanics or single particle quantum mechanics. 
%But it seems to lack a confident reasoning for the above equation. For what end should the Dirac-Frenkel principle be valid? 

%%%%%%%%%%%%%%%%%%
\subsection{Truncating a many-body state}

The limited or restricted forms of the state $|\Phi\rangle$ for the truncated many-body dynamics can in principle take any form. In the simplest case, assuming that the occupation numbers of bosons concentrate in one orbital for the whole time of evolution, we can treat the many-body state with one single-particle orbital. More generally, the state will reside in a sub-Hilbert space of a specific form.
An easily extendable and flexible form of the limitation on the size of the Hilbert space is the multiconfigurational time-dependent Hartree wavefunction ansatz, in which the many-body state of bosons is described as a linear combination of permanents $|\vec{n}\rangle$, with a finite number $M$ of orthonormalized time-dependent single-particle orbitals. Increasing the number $M$ of orbitals, we can easily extend the time-dependent sub-Hilbert space $\mathcal{M}(t)$. The basic steps in the procedure are as follows.
% As a result of introduction of time-varying orbitals or truncation, inverse of truncated SPDM comes in the evolution of orbitals. 
% As mentioned before, this hampers dynamical switching between various $M$ (the number of orbitals) values in Dirac-Frenkel's variational approach. 

Firstly, the many-body Hamiltonian %which specifies the energy and time evolution of the many-body state 
is given by
\be \label{H in field form}
\begin{split}
	&\hat{H} =
	\int \mathrm{d} \vec{x} ~~
	\hat{\Psi}^{\dag} (\vec{x})
	\bigg[
	-
	\frac{\nabla^2}
	{2m}
	+ 
	V_{\rm trap}(\vec{x})
	\bigg]
	\hat{\Psi} (\vec{x}) \\
	&+ 
	\frac{1}{2}
	\iint \mathrm{d} \vec{x}_{\alpha} \mathrm{d} \vec{x}_{\beta} \,
	\hat{\Psi}^{\dag} (\vec{x}_{\alpha})
	\hat{\Psi}^{\dag} (\vec{x}_{\beta})
	V(\vec{x}_{\alpha}, \vec{x}_{\beta})
	\hat{\Psi} (\vec{x}_{\beta})
	\hat{\Psi} (\vec{x}_{\alpha}).
\end{split}
\ee
With a complete set of basis orbitals, the field operators of creation and annihilation of particles are expressed by the expansions 
\be \label{fieldopexp}
\begin{split}
	\hat{\Psi}^{\dag}(\vec{x}) 
	= 
	\sum_{i=1}^{\infty}\hat{a}_i^{\dag}
	\phi_i^* (\vec{x})  ~
	\quad\textrm{and}\quad
	\hat{\Psi}(\vec{x},t) 
	= 
	\sum_{i=1}^{\infty}\hat{a}_i
	\phi_i (\vec{x},t) .
\end{split}
\ee
Using these, the Hamiltonian can be written as
\be \label{H in orbital form}
	\hat{H} 
	= \sum_{i,j=1}^{\infty}{ 
	\epsilon_{ij} 
	\hat{a}_{i}^{\dag} 
	\hat{a}_{j} }
	+ \frac{1}{2} 
	\sum_{i,j,k,l=1}^{\infty}{ 
	V_{ijkl}
	\hat{a}_{i}^{\dag} 
	\hat{a}_{j}^{\dag} 
	\hat{a}_{k} 
	\hat{a}_{l} }
\ee
where the single-particle matrix elements are
\be
	\epsilon_{ij} 
	= \int \mathrm{d} \vec{x} ~~
	\phi_i^* (\vec{x})
	\bigg[
	-\frac{\nabla^2}{2m}
	+ 
	V_{\rm trap} (\vec{x})
	\bigg]
	\phi_j (\vec{x}) ,
\ee
while the two-body interaction elements are represented by 
\be
	V_{ijkl}
	= \iint \mathrm{d} \vec{x}_{\alpha} \mathrm{d} \vec{x}_{\beta} ~
	\phi_i^* (\vec{x}_\alpha)
	\phi_j^* (\vec{x}_\beta)
	V(\vec{x}_\alpha, \vec{x}_\beta)
	\phi_k (\vec{x}_\beta)
	\phi_l (\vec{x}_\alpha) .
\ee
We abbreviate sometimes, for the sake of convenience, 
\be
\begin{split}
	&\hat{a}_{ij}^{\dag}
	\equiv
	\hat{a}_{i}^{\dag}
	\hat{a}_{j}, 
	\quad %%
	\hat{a}_{ijkl}^{\dag\dag}
	\equiv
	\hat{a}_{i}^{\dag}
	\hat{a}_{j}^{\dag}
	\hat{a}_{k}
	\hat{a}_{l}, \\
	&\hat{a}_{ijk}^{\dag\dag}
	\equiv
	\hat{a}_{i}^{\dag}
	\hat{a}_{j}^{\dag}
	\hat{a}_{k}, 
	\quad %%
	\hat{a}_{ijk}^{\dag}
	\equiv
	\hat{a}_{i}^{\dag}
	\hat{a}_{j}
	\hat{a}_{k} .
\end{split}
\ee
Then, the Hamiltonian can be written in short-hand form as
	$\hat{H} 
	= \hat{h} + \frac{1}{2}\hat{V}
	= \sum_{i,j=1}^{\infty}
	\epsilon_{ij} 
	\hat{a}_{ij}^{\dag}
	+ \frac{1}{2} 
	\sum_{i,j,k,l=1}^{\infty}
	V_{ijkl}
	\hat{a}_{ijkl}^{\dag\dag}$.

To find the many-body ground state $|G\rangle$, we have to minimize the energy expectation value $E_G = \langle G|\hat{H}|G\rangle$. Without any restrictions on the state $|G\rangle$, the actual exact ground state will be found. However, we cannot describe a state exactly as this will in general require infinitely many orbitals (or variables). So we confine the state into a space of finite dimension, i.e.
using only finite number $M$ of orbitals, which is computationally feasible, 
\be
	| G^{\mathcal{M}} \rangle =
	\sum_{\vec{n} \in \mathcal{M}} C_{\vec{n}} | \vec{n} \rangle.
\ee
The above is regarded as the truncation of the many-body bosonic state. Here $|\vec{n} \in \mathcal{M}\rangle$ indicates a normalized Fock state or a permanent
\be
	|\vec{n} \rangle
	\equiv
	\frac{
	\big(\hat{a}_1^{\dag}\big)^{n_1}
	\big(\hat{a}_2^{\dag}\big)^{n_2}
	\cdots
	\big(\hat{a}_M^{\dag}\big)^{n_M}}
	{\sqrt{n_1! n_2! \cdots n_M!}}
	|\textrm{vac} \rangle
	\quad\text{with}\quad
	\sum_{i=1}^{M} n_i = N,
\ee
which is a $N$ particle state of which the individual members are composed of $n_1$ particles in the $\phi_1 (\vec{x})$ orbital, $n_2$ particles in the $\phi_2 (\vec{x})$ orbital, $\cdots$, and $n_M$ particles in the $\phi_M (\vec{x})$ orbital. The $M$ orbitals must be orthonormalized to each other, 
\be
	\int \mathrm{d} \vec{x} ~ 
	\phi_i^* (\vec{x}) 
	\phi_j (\vec{x}) 
	= 
	\delta_{ij}.
\ee
These orbitals compose the sub-Hilbert manifold spanned by $M$ orthonormal orbitals, which will be denoted by $\mathcal{M}$ in our context. The operators of creation and annihilation on these are related to the field operators in position space by 
the inversion of Eq.\,\eqref{fieldopexp}, 
\be
	\hat{a}_i^{\dag}
	= 
	\int \mathrm{d} \vec{x} ~
	\hat{\Psi}^{\dag} (\vec{x}) \phi_i (\vec{x})
	\quad\textrm{and}\quad %%
	\hat{a}_i
	= 
	\int \mathrm{d} \vec{x} ~
	\hat{\Psi} (\vec{x})\phi_i^* (\vec{x}) .
\ee
%These primitive steps are well explained in \cite{MCHB, MCTDHB}. See those for a detailed account. \\
Here, a sub-Hilbert space spanned by $\sum_{i=1}^{M} c_{i} \phi_{i} (\vec{x})$ is to be chosen so as to describe the ground state optimally. Furthermore, the coefficients $C_{\vec{n}}$ which give the minimum of energy are to be determined.
Then,  $|G^{\mathcal{M}}\rangle$ can be considered as an optimal truncation of the actual ground state $|G\rangle$. The details on how to proceed concretely will follow below in section \ref{sec: Minimizing the energy}.

In the time-evolving case, we change not only the coefficients $C_{\vec{n}}$ along time, but also the sub-Hilbert space $\mathcal{M}$ changes with time. With a time-varying truncation of the many-body state, we can express the state as
\be \label{TD truncated many-body state}
	| \Phi (t) \rangle =
	\sum_{\vec{n} \in \mathcal{M}(t)} C_{\vec{n}} (t) | \vec{n} ; t \rangle ,
\ee
with time-varying orbitals and their conjugate creation operators
\be
	\hat{a}_i^{\dag} (t)
	= 
	\int \mathrm{d} \vec{x}~ \hat{\Psi}^{\dag} (\vec{x}) 
	\phi_i (\vec{x}, t) .
\ee
This approach was also incorporated in MCTDHB \cite{MCTDHB}. The time differentiation of the state $| \Phi (t) \rangle$ then becomes
\be \label{time differentiation of the state}
\begin{split}
	i \partial_t | \Phi (t) \rangle
	&=
	\sum_{\vec{n} \in \mathcal{M}(t)}
	\Big[
	\big( i \partial_t C_{\vec{n}} (t) \big)
	| \vec{n} ; t \rangle \\
	&+
	C_{\vec{n}} (t)
	\sum_{i=1}^{M}
	\int \mathrm{d} \vec{x} ~
	\big( i \partial_t \phi_i (\vec{x}, t) \big)
	\hat{\Psi}^{\dag} (\vec{x})
	\hat{a}_{i}
	| \vec{n} ; t \rangle
	\Big].
\end{split}
\ee
Expanding $i \partial_t \phi_i (\vec{x}, t)$ with a complete basis yields 
\be \label{t_ij_def}
	i \partial_t \phi_i (\vec{x}, t)
	=
	\sum_{k=1}^{\infty}
	\phi_{k} (\vec{x}, t) ~
	t_{ki} (t) .
\ee
where the matrix $t_{ki}$ for $1 \leq k \leq M$ indicates an {\em inner rotation} inside the sub-Hilbert space, whereas $t_{ki}$ for $k > M$ changes the sub-Hilbert space itself. Integrating both sides after multiplying with $\phi_{k}^{*} (\vec{x},t)$, we obtain
\be \label{orbital rotation t}
	t_{ki} (t)
	=
	\int \mathrm{d}\vec{x} ~
	\phi_{k}^{*} (\vec{x},t)
	i \partial_t \phi_i (\vec{x},t) .
\ee
Then Eq.\,(\ref{time differentiation of the state}) can be expressed in the alternative form
\be
\begin{split}
	i \partial_t | \Phi (t) \rangle
	=
	\sum_{\vec{n} \in \mathcal{M}(t)}
	\Big[
	&\big( i \partial_t C_{\vec{n}} (t) \big)
	| \vec{n} ; t \rangle \\
	&+
	C_{\vec{n}} (t)
	\sum_{k=1}^{\infty}
	\sum_{i=1}^{M}
	t_{ki}
	\hat{a}_{ki}^{\dag}
	| \vec{n} ; t \rangle
	\Big].
\end{split}
\ee
These preliminaries will be used in the following sections, providing the tools for describing the 
truncated state $|\Phi\rangle$ optimally. 
%% Since we do not count orbitals outside $M$, $\phi_{k}$'s for $k > M$ are just pictorial. The actual calculation takes upto $M$.

%Thus show nonequivalence of Dirac-Frenkel and MacLachlan

%Uniqueness of the solution 

%%%%%%%%%%%%%%%%%%
\subsection{Adapting the number of orbitals}
\label{subsec: Adapting the number of orbitals}

The governing equation of MCTDHB which comes from either Dirac-Frenkel's principle or TDVP implies that the SPDM must be always invertible, not only initially but also at any instant time afterwards. We quote here the equation (26) in \cite{MCTDHB}, which is 
\be
\begin{split}
	&i \partial_t \phi_{k} (\vec{x},t) \\
	&=
	\sum_{l=M+1}^{\infty}
	\phi_{l} (\vec{x},t)
	\bigg[
	\epsilon_{lk}
	+
	\sum_{i,n,p,q=1}^{M}
	V_{lnpq}
	\langle \rho \rangle^{-1}_{ki}
	\langle \hat{a}_{inpq}^{\dag\dag} \rangle
	\bigg]
\end{split}
\ee
in our notation, where $\langle \rho \rangle^{-1}_{ki}$ represents the inverse of the SPDM $\langle \rho_{ij} \rangle \equiv \langle \hat{a}_{ij}^{\dag} \rangle$. In \cite{MCTDHB}, it appears to be taken for granted that the SPDM is always invertible. When the SPDM becomes noninvertible, however, the MCTDHB method in fact suddenly fails. %Isn't this inevitable? Furthermore, why in the first place, with MCTDHB, should we be unable to propagate pure %single condensate into fragmented state? What was wrong? or what was missing?
%When the reduced SPDM $\langle \hat{a}_{ij}^{\dag} \rangle$ becomes noninvertible, MCTDHB which comes from %Dirac-Frenkel's principle fails. 
For example, when the whole bosons of our interest reside initially in a single orbital, MCTDHB cannot propagate this pure condensate into any fragmented state. The method does not provide a direct way to find the form of the second orbital in this case.

To decrease the sub-Hilbert space or the number of orbitals is not an issue. We can simply eliminate those orbitals with an occupation number ignorable, i.e.\,\,not of order $N$. However, to increase the sub-Hilbert space dimension, that is the number of orbitals $M$, becomes difficult since an increase of the sub-Hilbert space and an additional orbital must be set up optimally. 
Noninvertibility of the SPDM can obviously also happen dynamically. Unoccupied orbitals or scarcely occupied orbitals thus cause a problem with the dynamics of MCTDHB. These problems cannot be resolved by Dirac-Frenkel's principle or the TDVP. As McLachlan's principle is based on requiring that the least error should be acquired during time evolution, we may resolve this problem by finding an additional orbital which minimizes the error, as will be expounded in detail below.
%Applying McLachlan's alternatively, we have tried to solve these problems and to see things in a different point of view. 

%noninvertibility of single-particle density matrix 
%during time evolution hampers dynamical switching between various $M$ values for Dirac-Frenkel
%Explain why not a problem for MacLachlan

McLachlan's principle does not use the action, but the concept of error or remainder. We know the exact form of 
the many-body Schr\"odinger equation. It is
	$i \partial_t | \Phi \rangle
	= \hat{H} | \Phi \rangle$ with the full Hamiltonian Eq.\,(\ref{H in field form}) in second quantization form. Since the exact calculation is too cumbersome, we can represent the state by the relatively simple multiconfigurational time-dependent Hartree form. The remainder from exact evolution in any case becomes
\be
	\big[
	i \partial_t 
	- \hat{H}
	\big] 
	\big| \Phi \big\rangle .
\ee
In this expression, the left part $i \partial_t | \Phi \big\rangle$ is an evolution of the state $|\Phi\rangle$ in its limited, truncated form and the right part $\hat{H} | \Phi \big\rangle$ represents the exact evolution. Here, the initial state is specified in its truncated form, while the Hamiltonian $\hat{H}$ is not truncated.

A quantitative measure of the instantaneous error is then
\be \label{instantaneous error 0}
	\big\langle \Phi \big|
	\big[
	i \partial_t 
	- \hat{H}
	\big]^{\dag}
	\big[
	i \partial_t
	- \hat{H}
	\big]
	\big| \Phi \big\rangle
	\geq 0 ,
\ee
which is (by definition) positive semidefinite. We have to minimize this error by varying the truncated state in Eq.\,(\ref{TD truncated many-body state}).

%%%%%%%%%%%%%%%%%%%%%%%%%%%
\section{Minimizing the energy of the truncated many-body state}
\label{sec: Minimizing the energy}
\subsection{Variational method with Lagrange multipliers}

Before investigating the dynamics, let us demonstrate the time-independent scheme first. To find the many-body ground state, we have to minimize the energy expectation value
\be \label{Energy of tentative ground state with M orbitals}
	E_{G^{\mathcal{M}}} 
	=
	\langle G^{\mathcal{M}}|
	\hat{H}
	|G^{\mathcal{M}}\rangle
	=
	\Big[
	\sum_{\vec{m} \in \mathcal{M}} \langle \vec{m} | C_{\vec{m}}^*
	\Big]
	\hat{H}
	\Big[
	\sum_{\vec{n} \in \mathcal{M}} C_{\vec{n}} | \vec{n} \rangle
	\Big]
\ee
by varying the coefficients $C_{\vec{n}}$'s and the set of $M$ orbitals \{$\phi_i$\}, subject to the $(1+M^2)$ constraints
\be \label{Constraint on coefficients}
	\sum_{\vec{n}} C_{\vec{n}}^* C_{\vec{n}} = 1
	\qquad \text{(1 constraint)}
\ee
and
\be \label{Constraints on orbitals}
\begin{split} 
	\int \mathrm{d} \vec{x} ~ \phi_i^* (\vec{x}) \phi_j (\vec{x}) = \delta_{ij}
	\quad
	\textrm{for }i, j \leq M 
	\\ \text{($M^2$ constraints)}.
\end{split}
\ee
The variation with respect to the expansion coefficients $C_{\vec{n}}$ gives
\be \label{Variation wrt C_n}
	\frac{\partial E_{G^{\mathcal{M}}}}
	{\partial C_{\vec{n}}^*}
	= 
	\langle \vec{n} |
	\hat{H}
	\sum_{\vec{m} \in \mathcal{M}}
	C_{\vec{m}} | \vec{m} \rangle
	= 
	\lambda C_{\vec{n}}
	=
	E_{G^{\mathcal{M}}} C_{\vec{n}}, 
\ee
where $\vec{n} \in M$ \cite{footnote1}. We have a real functional $E_{G^M}$ Eq.\,(\ref{Energy of tentative ground state with M orbitals}) to be minimized, and one real equation of constraint, Eq.\,(\ref{Constraint on coefficients}), with $\frac{(N+M-1)!}{N!(M-1)!}$ complex variables $C_{\vec{n}}$ when we fix the total number of bosons at $N$. The undetermined Lagrange multiplier $\lambda$ which has to be real is determined to be $E_{G^{\mathcal{M}}}$ with the help of the constraint in Eq.\,(\ref{Constraint on coefficients}). 

Using the properties $\epsilon_{ij}^* = \epsilon_{ji}$, $V_{ijkl} = V_{jilk}$ and $V_{ijkl}^* = V_{lkji} = V_{klij}$, we can express the above equation explicitly as 
\be \label{Eq of coefficients for the ground state}
\begin{split}
	&\langle \vec{n} | 
	\hat{H} 
	\sum_{\vec{m} \in \mathcal{M}}
	C_{\vec{m}}
	| \vec{m} \rangle 
	= 
	E_{G^{\mathcal{M}}} C_{\vec{n}} \\
	&= \sum_l
	\bigg[
	\epsilon_{ll}
	+ 
	\frac{1}{2} 
	V_{llll}
	(n_l-1)
	\\
	&~~~~~~~~~~~~~~~~~~~
	+ \frac{1}{2}
	\sum_{k}{}^{'}
	(V_{lklk} + V_{lkkl})
	n_k
	\bigg]
	n_l
	C_{\vec{n}} \\
	&~+ 
	\sum_{j,l}{}^{'}
	\bigg[
	\epsilon_{lj}
	+ V_{lllj}
	(n_l-1)
	+ V_{ljjj}
	n_j \\
	&~~~~~~~~~~~~~~~~~~~
	+ \sum_{k}{}^{'}
	(V_{lkkj} + V_{lkjk})
	n_k
	\bigg]
	\sqrt{(n_j+1) n_l}
	C_{\vec{n}^{j}_{l}} \\
	&~~+ \frac{1}{2}
	\sum_{j,l}{}^{'}
	V_{lljj}
	\sqrt{(n_j+2)(n_j+1)(n_l-1)n_l}
	C_{\vec{n}^{jj}_{ll}} \\
	&~~+ \frac{1}{2}
	\sum_{i,j,l}{}^{'}
	V_{llji}
	\sqrt{(n_i+1)(n_j+1)(n_l-1)n_l}
	C_{\vec{n}^{ij}_{ll}} \\
	&~~+ \frac{1}{2}
	\sum_{j,k,l}{}^{'}
	V_{lkjj}
	\sqrt{(n_j+2)(n_j+1)n_k n_l}
	C_{\vec{n}^{jj}_{kl}} 
	\\&~~
	+ \frac{1}{2}
	\sum_{i,j,k,l}{}^{'}
	V_{lkji}
	\sqrt{(n_i+1)(n_j+1)n_k n_l}
	C_{\vec{n}^{ij}_{kl}},
\end{split}
\ee
where the primed summation $\sum'$ is performed such that
%goes over only different values for different indices which are themselves different. It means that 
different indices can have only different values. For example, summations over two and three indices are 
$\sum_{j,l}{}^{'} \equiv \sum_{j} \sum_{l \neq j}$, $\sum_{i,j,l}{}^{'} \equiv \sum_{i} \sum_{j \neq i} \sum_{l \neq (i,j)}$, and $\sum_{k}{}^{'}$ inside a bracket means $\sum_{k \neq \text{(the other indices)}}$.
%\col{This is quite unclear to me, what you mean here. Is this the same, like e.g. $\sum_{j,k}^{'} \equiv \sum_{j\neq k}$ or does it mean something different? And what would then $\sum_k^{'}$ mean, i.e., if there is just one index to be summed over?}
Though Eq.\,\eqref{Eq of coefficients for the ground state} is just an eigenvalue equation with fixed matrix components, the terms $\epsilon_{ij}$ and $V_{ijkl}$ are matrix elements depending on the orbitals.

For variation with respect to the orbitals $\{\phi_k\}$, functional differentiation is used. As the permanent $| \vec{n} \rangle$ is constructed from repeatedly applying creation operators of particles in the $M$ orbitals, it can be regarded to be given by multiple integrations over the orbitals $\phi_k (\vec{x})$, 
\be
	|\vec{n}\rangle
	=
	\int \mathrm{d}\vec{x}_{\alpha}
	\phi_{i} (\vec{x}_{\alpha})
	\hat{\Psi}^{\dag} (\vec{x}_{\alpha})
	\int \mathrm{d}\vec{x}_{\beta}
	\phi_{j} (\vec{x}_{\beta})
	\hat{\Psi}^{\dag} (\vec{x}_{\beta})
	\cdots
	|\textrm{vac}\rangle .
\ee
As each functional differentiation contributes to the result, this will be counted by a factor $\hat{a}^{\dag}_{k}$, which results in 
	$\frac{\partial \langle \Phi |}
	{\partial \phi_k^* (\vec{x})} 
	= \langle \Phi | 
	\hat{a}_{k}^{\dag} 
	\hat{\Psi} (\vec{x})$. But as the full Hamiltonian $\hat{H}$, which is given in field form by Eq.(\ref{H in field form}), is independent on the $M$ orbitals chosen, the functional differentiation of $\hat{H}$ with respect to the orbitals $\{\phi_k\}$ gives zero, 
	$\frac{\partial \hat{H}}
	{\partial \phi_k^* (\vec{x})}
	= 0$. Then functional variation of $E_{G^{\mathcal{M}}}$ in Eq.\,(\ref{Energy of tentative ground state with M orbitals}) with respect to the orbitals $\{\phi_k\}$, combined with the functional constraints Eq.\,(\ref{Constraints on orbitals}), leads to 
\be \label{Variation wrt phi_k}
\begin{split}
	&\frac{\partial E_{G^{\mathcal{M}}}}{\partial \phi_k^* (\vec{x})}
	= \\
	&\Big[
	\sum_{\vec{m} \in \mathcal{M}}
	\langle \vec{m} | C_{\vec{m}}^*
	\Big]
	\hat{a}_k^{\dag}
	\hat{\Psi} (\vec{x})
	\hat{H} 
	\Big[
	\sum_{\vec{n} \in \mathcal{M}}
	C_{\vec{n}} | \vec{n} \rangle
	\Big]
	= \sum_{j=1}^{M} \lambda_{kj} \phi_j (\vec{x}) ,
\end{split}
\ee
where $\lambda_{jk} = \lambda_{kj}^*$ is a Hermitian matrix. Here the method of Lagrange multipliers with complex functional variables is used \cite{footnote1}. Integrating each side over space after multiplication with $\phi_{l}^* (\vec{x})$ yields 
\be
	\Big[
	\sum_{\vec{m} \in \mathcal{M}}
	\langle \vec{m} |
	C_{\vec{m}}^*
	\Big]
	\hat{a}_k^{\dag}
	\hat{a}_l
	\hat{H}
	\Big[
	\sum_{\vec{n} \in \mathcal{M}}
	C_{\vec{n}}
	| \vec{n} \rangle 
	\Big]
	= 
	\begin{cases}
	\lambda_{kl} & \textrm{for } l \leq M \\
	0 & \textrm{for } l > M .
	\end{cases}
\ee
	Here, using Eq.\,(\ref{Variation wrt C_n}) and the property that $\vec{m}_{k}^{l} \in \mathcal{M}$ when $k,l \leq M$ and $\vec{m} \in \mathcal{M}$ \cite{footnote2}, 
	the undetermined set of Lagrange multipliers $\lambda_{kl}$ becomes related to $E_{G^{\mathcal{M}}}$ by $\lambda_{kl} = E_{G^{\mathcal{M}}} \langle \hat{a}_{kl}^{\dag} \rangle$ for $k, l \leq M$.
	Using Eq.\,(\ref{H in field form}) and the bosonic commutation relations between field operators, i.e. 
	$[\hat{\Psi}(\vec{x}), 
	\hat{\Psi}^{\dag}(\vec{x}')]
	=
	\delta
	\big(
	\vec{x} - \vec{x}'
	\big)$, 
	$[\hat{\Psi}(\vec{x}), 
	\hat{a}_{k}^{\dag}]
	=
	\phi_k (\vec{x})$, and so on, Eq.\,(\ref{Variation wrt phi_k}),  
	$\langle
	\hat{a}_k^{\dag}
	\hat{\Psi} (\vec{x})
	\hat{H}
	\rangle
	=
	\sum_{j=1}^{M} \lambda_{kj} \phi_j (\vec{x})$,  
	is explicitly expressed as 
\be
\begin{split}
	&\sum_{j=1}^{\infty}
	\langle
	\hat{a}_{kj}^{\dag}
	\rangle
	\bigg[
	-
	\frac{ \nabla^2}
	{2m}
	+ 
	V_{\rm trap}(\vec{x})
	\bigg]
	\phi_j (\vec{x}) \\
	&+
	\sum_{p,q,j=1}^{\infty}
	\langle
	\hat{a}_{kpqj}^{\dag\dag}
	\rangle
	\int \mathrm{d} \vec{x}' ~
	\phi_p^* (\vec{x}')
	V(\vec{x}, \vec{x}')
	\phi_q (\vec{x}')
	\phi_j (\vec{x}) \\
	&+ 
	\langle
	\hat{a}_k^{\dag}
	\hat{H}
	\hat{\Psi} (\vec{x})
	\rangle
	=
	\sum_{j=1}^{M} \lambda_{kj} \phi_j (\vec{x}) .
\end{split}
\ee
Since we can eliminate the annihilations above $M$, we obtain 
\be \label{Eq of orbitals for the ground state}
\begin{split}
	&\sum_{j=1}^{M}
	\langle
	\hat{a}_{kj}^{\dag}
	\rangle
	\hat{h}
	\phi_j (\vec{x})
	+ \sum_{p,q,j=1}^{M}
	\langle
	\hat{a}_{kpqj}^{\dag\dag}
	\rangle
	\hat{V}_{pq}
	\phi_j (\vec{x}) \\
	&= 
	\sum_{j=1}^{M} 
	\Big[
	\lambda_{kj} 
	-
	\langle
	\hat{a}_k^{\dag}
	\hat{H}
	\hat{a}_j
	\rangle
	\Big]
	\phi_j (\vec{x}) \\
	&= 
	\sum_{j=1}^{M} 
	\Big[
	E_{G^M}
	\langle
	\hat{a}_k^{\dag}
	\hat{a}_j
	\rangle
	-
	\langle
	\hat{a}_k^{\dag}
	\hat{H}
	\hat{a}_j
	\rangle
	\Big]
	\phi_j (\vec{x})
	\equiv 
	\sum_{j=1}^{M} 
	\tilde{\lambda}_{kj}
	\phi_j (\vec{x}),
\end{split}
\ee
where the new undetermined Lagrange multipliers
	$\tilde{\lambda}_{kj} 
	\equiv
	\lambda_{kj} 
	-
	\langle
	\hat{a}_k^{\dag}
	\hat{H}
	\hat{a}_j
	\rangle$ which satisfy 
	$\tilde{\lambda}_{kj}^*
	=
	\tilde{\lambda}_{jk}$, are introduced. We abbreviated, for the sake of convenience, two single-particle 
	and interaction operators defined by
\be
	\hat{h}
	\phi_j (\vec{x})
	\equiv
	\bigg[
	-
	\frac{\nabla^2}
	{2m}
	+ 
	V_{\rm trap}(\vec{x})
	\bigg]
	\phi_j (\vec{x})
\ee
and
\be
	\hat{V}_{pq}
	\phi_j (\vec{x})
	\equiv
	\int \mathrm{d} \vec{x}' ~
	\phi_p^* (\vec{x}')
	V (\vec{x}, \vec{x}')
	\phi_q (\vec{x}')
	\phi_j (\vec{x}) .
\ee 

%%%%%%%%%%%%%%%%%%%
\subsection{Method of steepest constrained descent}

To find the tentative ground state $| G^{M} \rangle$, we have to find the coefficients $C_{\vec{n}}$ and the complex orbital functions $\phi_k (\vec{x})$ satisfying Eq.\,(\ref{Eq of coefficients for the ground state}) and (\ref{Eq of orbitals for the ground state}) simultaneously. Additionally, the solutions must satisfy all $(1 + M^2)$ constraints Eq.\,(\ref{Constraint on coefficients},\,\ref{Constraints on orbitals}). The number of real values which we should find is $2\frac{(N+M-1)!}{N!(M-1)!}$ for the set of the $C_{\vec{n}}$, %actually infinite or 
$2M$ real functions for the $\phi_k (\vec{x})$, and $(1 + M^2)$ real values for the undetermined Lagrange multiplier $E_{G^M}$ and the $\lambda_{kj}$. The number of given real equations therefore is %in a form of real equating is
$2\frac{(N+M-1)!}{N!(M-1)!}$ for Eq.\,(\ref{Eq of coefficients for the ground state}), $2M$ real functional equations for Eq.\,(\ref{Eq of orbitals for the ground state}), and $(1 + M^2)$ for Eq.\,(\ref{Constraint on coefficients},\,\ref{Constraints on orbitals}).
Apart from the large number of variables, the equations (\ref{Constraint on coefficients},\,\ref{Constraints on orbitals},\,\ref{Eq of coefficients for the ground state},\,\ref{Eq of orbitals for the ground state}) are coupled. %entanged to each others and to each and every variables. 
To find a self-consistent solution is therefore obviously a very difficult problem. 
%is much more difficult problem than finding equations that the solution must satisfy. 
%So we have to know, in addition, the way to find $|G^M \rangle$, actually $C_{\vec{n}}$'s and $\phi_k (\vec{x})$'s from given equations. 

In ref.\,\cite{MCHB}, the authors started from an initial guess, then iteratively, 
% appropriate values again 
with a convergence check, they obtained a solution. A few years later in ref.\,\cite{MCTDHB}, applying the Wick rotation $it \rightarrow \tau$ on the equations of motion, they introduced the so-called imaginary time propagation. They stated that this reduces any arbitrary initial many-body state after a sufficient time 
of propagation to the ground state. 
The imaginary time evolution  
%\bdm
$	i \frac{\partial}
	{\partial t}
	|\Phi \rangle
	=
	\hat{H}
	|\Phi \rangle
	\Rightarrow 
	- 
	\frac{\partial}
	{\partial \tau}
	|\Phi \rangle
	=
	\hat{H}
	|\Phi \rangle
%\edm
$
implies that 
	$e^{- i \hat{H} t }
	\Rightarrow
	e^{- \hat{H} \tau } 
	|\Phi \rangle$.
 As $\tau$ goes to infinity, this seems to indicate that only the ground state survives and the excited states would no longer contribute. The contribution of the excited states decays exponentially according to a factor which is proportional to the energy difference from the ground state and to $\tau$. 
%(The reason why we introduce another method.) 
%No flexibility.
%Orbitals must be orthonormalized. But I think imageinary time propagation does not guarantee it. Although normalization of C_n can change, keeping orbitals' orthonormality might be safe.

Here, we present the method of steepest {\it constrained} descent, which is instrumental in constructing  the theory and design of higher-order algorithms of optimization {\it with constraints}. Since our problem requires optimization with {\it constraints on the variables} Eq.\,(\ref{Constraint on coefficients},\,\ref{Constraints on orbitals}), a simple-minded gradient (or, as it is more commonly termed steepest) descent method does not apply in our case. There are already numerous higher-order optimization algorithms {\it without constraints} such as Newton's method, conjugate gradient method, BFGS (Broyden-Fletcher-Goldfarb-Shanno) method, and the  Barzilai-Borwein method \cite{Keown}. But these methods are not applicable to our problem since the function to be minimized {\it with constraints on the variables} is generally {\it unbounded without constraints}. For example, the energy $\langle \hat{H} \rangle$ with unnormalized state $| \Phi \rangle$ can be zero, or $\pm\infty$. So when the actual ground-state energy is positive, the global minimum is where all $C_{\vec{n}} = 0$. For this case, e.g. Newton's method will simply lead to an incorrect solution. In other words, since our final point is not a global minimum without constraints, it can not be approximated as a quadratic function $\big[ f(\bf{x}) = \frac{1}{2} \bf{x}^{T} A \bf{x} + \bf{b}^{T} \bf{x} \big]$ around the minimum point (with constraints included). So all of these above-mentioned other methods, %(such as Newton's method, conjugate gradient method, BFGS method, and Barzilai-Borwein method) 
which are based on the expansion properties of the function around the minimum to second order will fail.
%Anyway the understanding of steepest {\it constrained} descent is fundamental to the theory and design of higher-order algorithms of optimization {\it with constraints}.

On the other hand, the method of steepest {\it constrained} descent guarantees that any given state is propagated to the neighboring lowest value point along the steepest {\it constrained} path for given variational parameters. Though it propagates a state only to a local neighboring minimum point, we can find in many cases the global minimum, or the ground state, from a well-chosen initial state, and with in addition well-chosen variational parameters. Although this does not deliver the state along the shortest path, the very large number of degrees of freedom on the choice of variational parameters or the sequential processing (separation) of variations can compensate in many cases. Furthermore the step size can be determined by one-point or two-point methods. The flexibility of the method of steepest constrained descent will therefore be beneficial for finding the ground state.

Let us see the process in more detail. The first step is to find (by educated guess) an appropriate initial state, specifying the coefficients $C_{\vec{n}}$ and the $M$ orbitals which we believe are appropriate to approximately describe the ground state of a given system. %It doesn't have to be accurate, though good guess would be definitely better to save time and calculation costs. 
From this initial guess, the state is propagated as follows. For the expansion coefficients $C_{\vec{n}}$, using the steepest constrained descent \cite{SCD}, 
\be
	\frac{d C_{\vec{n}}}
	{d \tau}
	=
	-
	\Delta_{C}(\tau)
	\bigg[
	\langle \vec{n} |
	\hat{H}
	\sum_{\vec{m}}
	C_{\vec{m}} | \vec{m} \rangle
	-
	\lambda
	C_{\vec{n}},
	\bigg]
\ee
where $\Delta_{C}(\tau)$ is any arbitrary positive function of $\tau$ which is introduced to satisfy 
	$\sum_{\vec{n}} 
	\frac{d C_{\vec{n}}^*}{d\tau}
	\frac{d C_{\vec{n}}}{d\tau}=$\,constant  
	at a certain given instant $\tau$ and therefore can be chosen in a convenient way to save time and calculation costs. Since it must satisfy $\sum_{\vec{n}} C_{\vec{n}}^* C_{\vec{n}} = 1$, i.e.
	$\sum_{\vec{n}}
	\big[
	C_{\vec{n}}^*
	\frac{d C_{\vec{n}}}
	{d \tau}
	+
	\frac{d C_{\vec{n}}^*}
	{d \tau}
	C_{\vec{n}}
	\big]
	=
	0$, $\lambda$ is to be $\langle \hat{H} \rangle$. 

As another option, we can use a polar representation for $C_{\vec{n}}$. Representing the complex variable $C_{\vec{n}}$ by an Euler representation with a radius $\xi_{\vec{n}}$ and an angle $\theta_{\vec{n}}$ gives 
	$C_{\vec{n}} 
	= \xi_{\vec{n}} 
	e^{i \theta_{\vec{n}}}$. 
	Then the constraint Eq.\,(\ref{Constraint on coefficients}) becomes 
	$\sum_{\vec{n}} \xi_{\vec{n}}^2 = 1$, restricting only the radial component of the complex variables  $C_{\vec{n}}$. Since we do not have to confine the variable change into the specific form 
	$\sum_{\vec{n}} 
	\big[
	\big(
	\frac{d \xi_{\vec{n}}}{d\tau}
	\big)^2
	+
	\xi_{\vec{n}}^2
	\big(
	\frac{d \theta_{\vec{n}}}{d\tau}
	\big)^2
	\big] =$\,constant, 
	separating the two variable sets can be much more efficient in this case. That is, we set 
	$\frac{d \xi_{\vec{n}}}{d\tau} = 0$ and 
	$\sum_{\vec{n}} 
	\big(
	\frac{d \theta_{\vec{n}}}{d\tau}
	\big)^2 =$\,const 
	for the selected $\tau$ time spans which would be chosen in a computationally favorable way, as well as 
	$\frac{d \theta_{\vec{n}}}{d\tau} = 0$ 
	and 
	$\sum_{\vec{n}} 
	\big(
	\frac{d \xi_{\vec{n}}}{d\tau}
	\big)^2 =$\,const for the other $\tau$ spans. 
	Then the method of steepest constrained descent gives
\be
	\frac{d \xi_{\vec{n}}}
	{d \tau} = 0 ,
	\qquad %%
	\frac{d \theta_{\vec{n}}}
	{d \tau}
	=
	-
	\Delta_{\theta}(\tau)
	\Im
	\big(
	\xi_{\vec{n}}
	e^{- i \theta_{\vec{n}}}
	\langle \vec{n} |
	\hat{H}
	\rangle 
	\big),
\ee
for some given $\tau$ spans, and
\be
	\frac{d \theta_{\vec{n}}}
	{d \tau}
	= 0 ,
	\qquad %%
	\frac{d \xi_{\vec{n}}}
	{d \tau}
	=
	-
	\Delta_{\xi}(\tau)
	\Big[
	\Re
	\big(
	e^{- i \theta_{\vec{n}}}
	\langle \vec{n} |
	\hat{H}
	\rangle
	\big)
	-
	\lambda
	\xi_{\vec{n}}
	\Big]
\ee
for the other following $\tau$-intervals. 
Here $\Re$ means real part of complex number and $\Im$ means imaginary part of complex number. Using $\sum_{\vec{n}} \xi_{\vec{n}}^2 = 1$, and therefore 
	$\sum_{\vec{n}} 
	2 
	\xi_{\vec{n}} 
	\frac{d \xi_{\vec{n}}}
	{d \tau}
	=
	0$, then 
	$\lambda 
	= \langle
	\hat{H}
	\rangle$. 
	Dealing with $\xi_{\vec{n}}$ and $\theta_{\vec{n}}$ separately, we propagate the two variable sets successively and  iteratively until convergence is achieved. Employing any sequence, they will ultimately approach the minimum.  %With good choice of variables and topology, it will approach the minimum more quickly. \\

For the orbitals $\phi_k (\vec{x})$, the method of steepest constrained descent gives
\be \label{orbital descent}
\begin{split}
	&\frac{d \phi_{k} (\vec{x})}
	{d \tau}
	= 
	-\Delta_{\phi_k} (\tau)
	\Big[
	\langle
	\hat{a}_{k}^{\dag}
	\hat{\Psi} (\vec{x})
	\hat{H}
	\rangle
	-
	\sum_{j=1}^{M}
	\lambda_{kj}
	\phi_{j} (\vec{x})
	\Big] \\
	&=
	-
	\Delta_{\phi_k} (\tau)
	\bigg[
	\sum_{j=1}^{M}
	\langle
	\hat{a}_{kj}^{\dag}
	\rangle
	\hat{h}
	\phi_j (\vec{x})
	+ \sum_{p,q,j=1}^{M}
	\langle
	\hat{a}_{kpqj}^{\dag\dag}
	\rangle
	\hat{V}_{pq}
	\phi_j (\vec{x}) \\
	&~~~~~~~~~~~~~~~~~~
	- 
	\sum_{j=1}^{M} 
	\tilde{\lambda}_{kj} 
	\phi_j (\vec{x})
	\bigg].
\end{split}
\ee
If we propagate the orbitals separately one after another, % iteratively, 
i.e. only the $k$-th orbital changes within a certain period, the constraint becomes 
	$\int \mathrm{d}\vec{x}~
	\phi_l^* (\vec{x})
	\frac{\phi_k (\vec{x})}{d\tau}
	= 0$ for $l \neq k$ and 
	$\int \mathrm{d}\vec{x}~
	\big(
	\phi_k^* (\vec{x})
	\frac{\phi_k (\vec{x})}{d\tau}
	+
	\frac{\phi_k^* (\vec{x})}{d\tau}
	\phi_k (\vec{x})
	\big)
	= 0$. 
	Then the undetermined Lagrange multipliers becomes
\be
	\tilde{\lambda}_{kl}
	=
	\sum_{j=1}^{M}
	\langle
	\hat{a}_{kj}^{\dag}
	\rangle
	\epsilon_{lj}
	+ \sum_{p,q,j=1}^{M}
	\langle
	\hat{a}_{kpqj}^{\dag\dag}
	\rangle
	V_{lpqj}
\ee
and
\be
	\tilde{\lambda}_{kk}
	=
	\Re \Big(
	\sum_{j=1}^{M}
	\langle
	\hat{a}_{kj}^{\dag}
	\rangle
	\epsilon_{kj}
	+ \sum_{p,q,j=1}^{M}
	\langle
	\hat{a}_{kpqj}^{\dag\dag}
	\rangle
	V_{kpqj}
	\Big) .
\ee
Separating the propagation of the orbitals, i.e. propagating the orbitals one after another independently, makes the process easier to control.  %And the computational error which comes from finite step size can be handled by cutting the non-orthogonal part of the only propagating orbital.

In a numerical implementation, the method of steepest {\it constrained} descent is performed in discrete steps. Then a projection onto the set of constraints must be employed at every discrete step. Using just a one-point step size method, we can use the Lagrange multipliers $\lambda_{kj}$ as a degree of freedom to save computational cost, since we do not need to consider the propagation outside of the constraints. As an example, we can set $\lambda_{kj}
	=\langle
	\hat{a}_k^{\dag}
	\hat{H}
	\hat{a}_j
	\rangle$ in Eq.\,(\ref{orbital descent}) so that we do not need to calculate the quantities 
	$\langle
	\hat{a}_k^{\dag}
	\hat{H}
	\hat{a}_j
	\rangle$. However, to ensure fast convergence, the optimal step size is determined with a two-point or even four-point method. The role of Lagrange multipliers is crucial, then, because we mix two gradients at different points. Because of the finite step size, the propagation outside of the constraints at one point can be the direction of the steepest {\it constrained} descent at another point. 
The task in a concrete implementation therefore is to combine the proper directions with appropriately chosen Lagrange multipliers at different points,  to optimize the speed of convergence.

%%%%%%%%%%%%%%%%%%%%%%%%%%%
\section{Control of truncated many-body evolution}
\subsection{Evaluating the error of truncated many-body evolution}

The instantaneous error is expressed by Eq.\,(\ref{instantaneous error 0}). Minimizing this error with a state change under the truncation Eq.\,(\ref{TD truncated many-body state}) gives us 
the appropriate many-body evolution. This offers, as a major benefit of the present approach, a definite value of the error, which indicates how accurately the truncated evolution describes the exact one. 

Explicitly expressing the error, we have 
\be \label{instantaneous error}
\begin{split}
	&\langle \Phi |
	\big[
	\hat{H} - i \partial_t
	\big]^{\dag}
	\big[
	\hat{H} - i \partial_t
	\big]
	| \Phi \rangle \\
	&=
	\sum_{\vec{n} \in \mathcal{M}(t)}
	\bigg[
	\langle \vec{n} |
	C_{\vec{n}}^*
	\hat{H}
	+ 
	\langle \vec{n} |
	\big(i
	\partial_t
	C_{\vec{n}}^*
	\big) \\
	&~~~~~~~~~~~~~~~~
	+ 
	\sum_{i=1}^{M}
	\langle \vec{n} |
	C_{\vec{n}}^*
	\hat{a}_{i}^{\dag}
	\int \mathrm{d} \vec{x}
	\big(i
	\partial_t
	\phi_i^* (\vec{x},t)
	\big)
	\hat{\Psi} (\vec{x})
	\bigg] \\
	&~
	\times 
	\sum_{\vec{m} \in \mathcal{M}(t)}
	\bigg[
	\hat{H}
	C_{\vec{m}}
	| \vec{m} \rangle
	-  
	\big(i
	\partial_t
	C_{\vec{m}}
	\big)
	| \vec{m} \rangle \\
	&~~~~~~~~~~~~~~~~
	-
	\sum_{j=1}^{M}
	\int \mathrm{d} \vec{x}'
	\big(i
	\partial_t
	\phi_j (\vec{x}',t)
	\big)
	\hat{\Psi}^{\dag} (\vec{x}')
	\hat{a}_{j}
	C_{\vec{m}}
	| \vec{m} \rangle
	\bigg] .
\end{split}
\ee
We minimize this instantaneous error, varying the complex variables 
	$\partial_t C_{\vec{n}} $ and 
	$\partial_t \phi_i (\vec{x},t) $ subject to the (1 + $M^2$) constraints
\be \label{coefficient constraint}
\begin{split}
	&\partial_t
	\bigg[
	\sum_{\vec{n}} C_{\vec{n}}^* (t) C_{\vec{n}} (t) = 1
	\bigg] \\
	&\Rightarrow \quad
	\sum_{\vec{n}} 
	\bigg[
	\big(
	\partial_t C_{\vec{n}}^* (t) 
	\big)
	C_{\vec{n}} (t) 
	+
	C_{\vec{n}}^* (t) 
	\big(
	\partial_t C_{\vec{n}} (t) 
	\big)
	\bigg]
	= 0 .
\end{split}
\ee
From the orthonormality condition, we obtain  
\be \label{orbital constraint}
\begin{split}
	&\partial_t
	\bigg[
	\int \mathrm{d} \vec{x}
	\phi_i^* (\vec{x},t)
	\phi_j (\vec{x},t)
	= \delta_{ij}
	\bigg]
	\Rightarrow \quad
	\int \mathrm{d} \vec{x}
	\bigg[
	\big(
	\partial_t \phi_i^* (\vec{x},t)
	\big) \\
	&\times
	\phi_j (\vec{x},t)
	+
	\phi_i^* (\vec{x},t)
	\big(
	\partial_t \phi_j (\vec{x},t)
	\big)
	\bigg]
	= 0 .
\end{split}
\ee
Eq.\,(\ref{coefficient constraint}) requires probability conservation of the state itself and Eq.\,(\ref{orbital constraint}) requires conservation of orthonormality of orbitals. Using the expression Eq.\,(\ref{orbital rotation t}), Eq.\,(\ref{orbital constraint}) can be expressed as the hermiticity condition $t_{ji}^* = t_{ij}$. 

Variation with respect to $\partial_t C_{\vec{n}}^* $ leads to
\be
	\langle \vec{n} | i
	\big[
	\hat{H} - i \partial_t
	\big]
	|\Phi\rangle
	=
	\lambda (t) C_{\vec{n}} ,
\ee
resulting in
\be
i \partial_t C_{\vec{n}}
	=
	\langle \vec{n} |
	\big[
	\hat{H} - 
	\sum_{i=1}^{\infty}
	\sum_{j=1}^{M}
	t_{ij} \hat{a}_{ij}^{\dag}
	\big]
	|\Phi\rangle
	+
	i \lambda (t) C_{\vec{n}}.
\ee
As the constraint Eq.\,(\ref{coefficient constraint}) enforces $\lambda (t) = 0$, 
\be \label{te_Eq1} % for the time evolution
	\langle \vec{n} |
	\big[
	\hat{H}
	-
	i \partial_t
	\big]
	|\Phi\rangle
	= 0;
\ee
hence the time evolution of the expansion coefficients takes the form
\be \label{te_Eq1-1}
\begin{split}
	i \partial_t
	C_{\vec{n}}
	=
	\langle \vec{n} |
	\big[
	\hat{H} - \hat{t}~
	\big]
	|\Phi\rangle
\end{split}
\ee
where $\vec{n} \in M(t)$ and $\hat{t} = \sum_{i,j} t_{ij} \hat{a}_{ij}^{\dag}$. 

Variation with respect to $\partial_t \phi_k^* (\vec{x},t)$ gives
\be
	\langle \Phi | i
	\hat{a}_{k}^{\dag}
	\hat{\Psi} (\vec{x})
	\big[
	\hat{H} - i \partial_t
	\big]
	|\Phi\rangle
	=
	\sum_{l=1}^{M}
	\lambda_{kl} \phi_{l} (\vec{x},t) .
\ee
After multiplying both sides with $\phi_l^* (\vec{x},t)$ and integrating, one finds
\be
	\langle \Phi | i
	\hat{a}_{k}^{\dag}
	\hat{a}_{l}
	\big[
	\hat{H} - i \partial_t
	\big]
	|\Phi\rangle
	=
	\begin{cases}
	\lambda_{kl} (t) &\text{ if } l \leq M \\ 
	0 &\text{ if } l>M . 
	\end{cases}
\ee
Since $\langle\vec{n}| \hat{a}_{kl}^{\dag}$ belongs to ${\mathcal M}(t)$ whenever $\vec{n} \in \mathcal{M}(t)$ and $k,l \leq M$, all $\lambda_{kl} (t)$ here become zero too with the help of Eq.\,(\ref{te_Eq1}). So for any $k$ and $l$,
\be \label{te_Eq2} % for the time evolution
	\langle\Phi |
	\hat{a}_{kl}^{\dag}
	\big[
	\hat{H} - i \partial_t
	\big]
	| \Phi \rangle
	= 0.
\ee
Expanding the above equation for $l>M$, 
\be
	\langle
	\hat{a}_{kl}^{\dag} \hat{H}
	\rangle
	=
	\sum_{j=1}^{M}
	\langle
	\hat{a}_{kj}^{\dag}
	\rangle
	\int \mathrm{d}\vec{x}
	\phi_l^* (\vec{x},t)
	\big(
	i \partial_t \phi_j (\vec{x},t)
	\big) .
\ee
Since the SPDM can be noninvertible, we reduce the density matrix by eliminating unoccupied orbitals in which the eigenvalues of the SPDM becomes zero (relative to $\ord(N)$ in the limit $N\rightarrow\infty$)
after diagonalizing the SPDM. In the process of diagonalizing the SPDM, the orbitals are unitarily transformed so that the essentially unoccupied orbitals can be found and eliminated. Introducing the inverse of this reduced density matrix 
	$\langle \rho_{ki} \rangle 
	\equiv 
	\langle \hat{a}_{ki}^{\dag} \rangle
	\equiv
	\langle \hat{a}_{k}^{\dag} \hat{a}_{i} \rangle$  
%\col{Why is there a dagger here on the $\hat a_{ki}$??}
when the $\ord(N)$ 
occupied orbitals exist up to the $M_1$th orbital ($\sum_{i=1}^{M_1} \langle \rho \rangle_{ki}^{-1} \langle \rho_{ij} \rangle = \delta_{kj}$) and using the completeness relation $\sum_{l=1}^{\infty} \phi_{l}^* (\vec{x}',t) \phi_{l} (\vec{x},t) = \delta (\vec{x}'-\vec{x})$, the evolution equation of the orbitals acquires the form
\begin{widetext}
\be \label{te_Eq2-1}
\begin{split}
	&i \partial_t
	\phi_k (\vec{x},t)
	=
	\sum_{l=1}^{M}
	t_{lk}
	\phi_{l} (\vec{x},t)
	+
	\sum_{l=M+1}^{\infty}
	\sum_{i=1}^{M_1}
	\langle
	\rho
	\rangle_{ki}^{-1}
	\langle
	\hat{a}_{i}^{\dag}
	\hat{a}_{l}
	\hat{H}
	\rangle
	\phi_{l} (\vec{x},t) \\
	&=
	\sum_{l=1}^{M}
	t_{lk}
	\phi_{l} (\vec{x},t)
	+
	\sum_{l=M+1}^{\infty}
	\sum_{i=1}^{M_1}
	\langle
	\rho
	\rangle_{ki}^{-1}
	\langle
	\bigg[
	\sum_{n=1}^{M_1}
	\epsilon_{ln}
	\hat{a}_{in}^{\dag}
	+
	\sum_{n,p,q=1}^{M_1}
	V_{lnpq}
	\hat{a}_{inpq}^{\dag\dag}
	\bigg]
	\rangle
	\phi_{l} (\vec{x},t) \\
	&=
	\sum_{l=1}^{M}
	t_{lk}
	\phi_{l} (\vec{x},t)
	+
	\sum_{l=M+1}^{\infty}
	\bigg[
	\epsilon_{lk}
	+
	\sum_{i,n,p,q=1}^{M_1}
	V_{lnpq}
	\langle
	\rho
	\rangle_{ki}^{-1}
	\langle
	\hat{a}_{inpq}^{\dag\dag}
	\rangle
	\bigg]
	\phi_{l} (\vec{x},t)
\end{split}
\ee
\end{widetext}
for $k \leq M_1$. 

We divide the evolution of the orbitals into two parts. We call the left $l\leq M$ part of Eq.\,(\ref{te_Eq2-1}) inner rotation and the right $l>M$ part of Eq.\,(\ref{te_Eq2-1}) rotation toward the outside the sub-Hilbert space. Since the sub-Hilbert space spanned by $M$ orbitals, $\sum_{i=1}^{M} c_{i} \phi_{i} (\vec{x},t)$,  does not change under inner rotation, we realize that only a rotation toward the outside deforms the sub-Hilbert space. For the evolution inside the sub-Hilbert space, i.e. the inner rotation, we simply use $t_{ij}$ defined in \eqref{t_ij_def}, 
which can be any Hermitian matrix. Using Eq.\,(\ref{te_Eq1},\,\ref{te_Eq2},\,\ref{te_Eq2-1}), the 
expression for the error Eq.\,(\ref{instantaneous error}) can be strongly simplified: 
\begin{widetext}
\be \label{instantaneous error 1}
\begin{split}
	&\langle \Phi |
	\big[
	\hat{H} - i \partial_t
	\big]^{\dag}
	\big[
	\hat{H} - i \partial_t
	\big]
	| \Phi \rangle
	=
	\langle \Phi |
	\big[
	\frac{1}{2}
	\sum_{i,j,k,l=1}^{\infty}
	V_{ijkl} \hat{a}_{ijkl}^{\dag\dag}
	\big]
	\big[
	\hat{H} - i \partial_t
	\big]
	| \Phi \rangle
	=
	\frac{1}{2}
	\sum_{i,j=1}^{M_1}
	\sum_{k,l=1}^{\infty}
	V_{ijkl} 
	\langle \Phi |
	\hat{a}_{ijkl}^{\dag\dag}
	\big[
	\hat{H} - i \partial_t
	\big]
	| \Phi \rangle \\
	&=
	-\!\sum_{i,j,k,n,r,s,p,q=1}^{M_1} \sum_{l=M+1}^{\infty}\!
	V_{ijkl} V_{lspq}
	\langle \hat{a}_{ijkn}^{\dag\dag} \rangle
	\langle \rho \rangle_{nr}^{-1}
	\langle \hat{a}_{rspq}^{\dag\dag} \rangle 
	+\!\sum_{i,j,k,s,p,q=1}^{M_1} \sum_{l=M+1}^{\infty}\!
	V_{ijkl} V_{lspq}
	\langle \hat{a}_{ijskpq}^{\dag\dag\dag} \rangle \\
	&~~~
	+\frac{1}{2}\!\sum_{i,j,p,q=1}^{M_1} \sum_{k=1}^{M} \sum_{l=M+1}^{\infty}\!
	V_{ijkl} V_{lkpq}
	\langle \hat{a}_{ijpq}^{\dag\dag} \rangle
	%\\&~~~
	+\frac{1}{2}
	\sum_{i,j,p,q=1}^{M_1} \sum_{k=1}^{\infty} \sum_{l=M+1}^{\infty}\!
	V_{ijkl} V_{lkpq}
	\langle \hat{a}_{ijpq}^{\dag\dag} \rangle .
\end{split}
\ee
\end{widetext}
The above equation represents our main result, rendering the error of many-body quantum evolution upon
truncation systematically computable. As clearly seen, the error stems entirely from interactions. In other words, the error does not depend on the choice of $t_{ij}$ and even on the single particle energy matrix $\epsilon_{ij}$ for any given truncated initial state, provided the evolution of the state,  $\partial_t |\Phi\rangle$, is optimally taken as in Eq.\,(\ref{te_Eq1},\,\ref{te_Eq2},\,\ref{te_Eq2-1}). 

%%%%%%%%%%%%%%%%%%%
\subsection{Determining the number of orbitals dynamically}

When the error becomes large or, alternatively, when we aim at describing the system more precisely, we have to increase the number of orbitals $M_1$ into $M$. The $(M - M_1)$ additional orbitals 
then can be determined by variation of the error with respect to $\phi_{u} (\vec{x},t)$ where $M_1 < u \leq M$, and subject to the orthonormalization condition 
	$\int \mathrm{d}\vec{x} ~
	\phi_{u}^{*} (\vec{x},t)
	\phi_{v} (\vec{x},t) 
	= \delta_{uv}$. The method of Lagrange multipliers for functional variables gives the stationarity condition
\begin{widetext} 
\begin{multline}
	\sum_{i,j,k,n,r,s,p,q=1}^{M_1}
	V_{ijku} \big( \hat{V}_{sp} \phi_{q} (\vec{x},t) \big)
	\langle \hat{a}_{ijkn}^{\dag\dag} \rangle
	\langle \rho \rangle_{nr}^{-1}
	\langle \hat{a}_{rspq}^{\dag\dag} \rangle 
	-
	\sum_{i,j,p,q=1}^{M_1} \sum_{k,s=1}^{M}
	V_{ijku} \big( \hat{V}_{sp} \phi_{q} (\vec{x},t) \big)
	\langle \hat{a}_{ijk}^{\dag\dag} \hat{a}_{spq}^{\dag} \rangle \\
	=
	\sum_{v=1}^{M}
	\mu_{uv} \phi_{v} (\vec{x},t) . \label{Eq for additional orbitals}
\end{multline}
\end{widetext} 
While this is an equation for the additional orbitals, the additional orbitals are difficult to obtain directly from the above equation. In pratice, we can choose at best some orbitals that will approximately satisfy the above equation.

Therefore, we use the method of steepest constrained descent again. From an initial trial orbital, we propagate the orbital toward
\be \label{additional orbital}
\begin{split}
	&\frac{d \phi_u (\vec{x})}{d \tau}
	=
	-\Delta_{\phi_u} (\tau)
	\bigg[
	\sum_{i,j,k,n,r,s,p,q=1}^{M_1}
	V_{ijku} \big( \hat{V}_{sp} \phi_{q} (\vec{x},t) \big) \\
	&\times
	\langle \hat{a}_{ijkn}^{\dag\dag} \rangle
	\langle \rho \rangle_{nr}^{-1}
	\langle \hat{a}_{rspq}^{\dag\dag} \rangle
	-
	\sum_{i,j,p,q=1}^{M_1} \sum_{k,s=1}^{M}
	V_{ijku} \big( \hat{V}_{sp} \phi_{q} (\vec{x},t) \big) \\
	&\langle \hat{a}_{ijk}^{\dag\dag} \hat{a}_{spq}^{\dag} \rangle
	-
	\sum_{v=1}^{M}
	\mu_{uv} \phi_{v} (\vec{x},t)
	\bigg],
\end{split}
\ee
so that the error become minimized with this additional orbital. If we propagate orbitals separately and iteratively one after another, i.e. only the $u$th orbital changes along $\tau$, the constraint becomes 
	$\int \mathrm{d}\vec{x}~
	\phi_v^* (\vec{x})
	\frac{\phi_u (\vec{x})}{d\tau}
	= 0$ for $v \neq u$ and 
	$\int \mathrm{d}\vec{x}~
	\big(
	\phi_u^* (\vec{x})
	\frac{\phi_u (\vec{x})}{d\tau}
	+
	\frac{\phi_u^* (\vec{x})}{d\tau}
	\phi_u (\vec{x})
	\big)
	= 0$. 
Then the undetermined Lagrange multipliers become
\be
\begin{split}
	\mu_{uv}
	=
	&\sum_{i,j,k,n,r,s,p,q=1}^{M_1}
	V_{ijku} V_{vspq}
	\langle \hat{a}_{ijkn}^{\dag\dag} \rangle
	\langle \rho \rangle_{nr}^{-1}
	\langle \hat{a}_{rspq}^{\dag\dag} \rangle 
	\\&
	-
	\sum_{i,j,p,q=1}^{M_1} \sum_{k,s=1}^{M}
	V_{ijku} V_{vspq}
	\langle \hat{a}_{ijk}^{\dag\dag} \hat{a}_{spq}^{\dag} \rangle ,
\end{split}
\ee
where the orbital index ranges are constrained to be $M_1 < u \leq M$ and $1 \leq v \leq M$. 
%Explicit expression for the error in terms of correlation functions/orbitals/expansion coefficients?

Since the inner rotation can be arbitrarily chosen, we can take $t_{ij}=\epsilon_{ij}$, which renders the result in a simple form. 
The evolution Eq.\,(\ref{te_Eq1-1}) for the expansion coefficients becomes
\be \label{te1}
	i \partial_t C_{\vec{n}}
	=
	\left\langle \vec{n} \right|
	\frac{1}{2}
	\sum_{i,j=1}^{M}
	\sum_{k,l=1}^{M_1}
	V_{ijkl}
	\hat{a}_{ijkl}^{\dag\dag}
	\left| \Phi \right\rangle .
\ee
This implies the desired property that, when the interaction is turned off, the coefficients $C_{\vec{n}}$ 
do not change at all. The Schr\"odinger equation for the orbitals, Eq.\,(\ref{te_Eq2-1}), becomes
\be \label{te2}
\begin{split}
	&i \partial_t \phi_k (\vec{x},t)
	=
	\hat{h} \phi_k (\vec{x},t) \\
	&+
	\sum_{l=M+1}^{\infty}
	\sum_{i,n,p,q=1}^{M_1}
	V_{lnpq}
	\langle \rho \rangle_{ki}^{-1}
	\langle \hat{a}_{inpq}^{\dag\dag} \rangle
	\phi_l (\vec{x},t)
\end{split}
\ee
for $k \leq M_1$. The projection toward the outside of the sub-Hilbert space $\mathcal{M}$ takes place only on the interaction term. Thus the option of taking $t_{ij}=\epsilon_{ij}$ for the inner rotation shows the effect of the interaction term in this explicit manner. 

For $M_1 < k \leq M$, the evolution of the additional orbitals can be chosen in any convenient way, since the time derivatives of the additional orbitals do not occur in the error measure. 
The only constraint is the inner rotation. As we have chosen 
	$t_{lk} 
	\big( 
	= \int \mathrm{d}\vec{x}
	\phi_l^* (\vec{x},t)
	i \partial_t \phi_k (\vec{x},t)
	\big)$ 
	for $M_1 < l \leq M$ and $k \leq M_1$ 
	to be $\epsilon_{lk}$,  it is a Hermitian matrix given by 
	$t_{lk} = t^*_{kl} = epsilon^*_{kl} = \epsilon_{lk}$, 
	satisfying the orthonormality constraint
	Eq.\,(\ref{orbital constraint}). 
Taking the rotation toward the outside of the sub-Hilbert space to be also equal to the single-particle energy matrix,  $\epsilon_{kl}$ for $k>M$, the evolution of the additional orbital $\phi_l (\vec{x},t)$ for $M_1<l\leq M$ 
is determined by the simple equation 
\be \label{te3}
	i \partial_t \phi_l (\vec{x},t)
	=
	\hat{h} \phi_l (\vec{x},t),
\ee
while the additional orbitals are found from Eq.\,(\ref{additional orbital}). 

We, finally, emphasize again that the key difference to the MCTDHB approach is that additional 
macroscopically occupied orbitals during dynamical evolution can be found in our approach. 
By this means, we can handle the exceptional case when the SPDM is not invertible, and increase the number of orbitals under any given circumstances and boundary conditions.

%%%%%%%%%%%%%%%%%%
\section{Summary}

Using McLachlan's principle and the methods of Lagrange multipliers and steepest constrained descent, we have developed a systematic method to describe the many-body evolution of bosons in a rigorously controlled manner. Writing the many-body state in Hartree form and limiting the size of the Hilbert space by truncating into a finite number of macroscopically occupied field operator modes, the error from the exact evolution can be minimized self-consistently. This gives a variationally optimized solution to the evolution of the truncated many-body state.

We have demonstrated that without two-body interactions, our scheme possesses the desired property 
that the evolution of the many-body state can be exactly described with zero error, cf.\,\,Eq.\,(\ref{instantaneous error 1}). 
When interactions are turned on, the error accumulates during time evolution. Employing our method, we can evolve the truncated many-body state in an optimized way. Monitoring the error, we can increase the accuracy of the evolution by increasing the number of orbitals in a self-consistent way. 
By adaptively changing the number of orbitals based on the instantaneous error measure, we can essentially automatically ensure the validity of the result for the many-body state.

We conclude by a brief summary of our approach when it is applied to the well-known and ubiquitous Gross-Pitaevski\v\i\/ equation.
We start by evolving the initial trial 
state $|\Phi\rangle$ along the $M=1$ version of Eq.\,(\ref{te1},\,\ref{te2}) and simultaneously check whether the error Eq.\,(\ref{instantaneous error 1}) remains small or not. Monitoring the error Eq.\,(\ref{instantaneous error 1}), we can determine under which conditions the Gross-Pitaevski\v\i\/ equation loses its validity. When this happens,  the error becoming large, we increase the number of orbitals to $M=2$ (thus, here, $M_1=1$ and $M=2$). The additional second orbital is found by the method of steepest constrained descent, using Eq.\,(\ref{additional orbital}). Then the subsequent evolution of the quantum many-body state follows the $M=2$ version of Eq.\,(\ref{te1},\,\ref{te2}), while the evolution of the (initially singular) second orbital follows Eq.\,(\ref{te3}). We monitor the error Eq.\,(\ref{instantaneous error 1}) again, checking that the error is decreased to a sufficient degree. In a self-consistent manner one then proceeds until some prescribed accuracy is obtained.

%\bigskip\noindent {\bf Acknowledgments}
\acknowledgments
%\medskip 

This research was supported by the Brain Korea BK21
program and the NRF of Korea, Grant Nos. 2010-0013103, 2011-0029541, and 2014R1A2A2A01006535.
%2014-006535. 

%\end{Acknowledgements}

%%%%%%%%%%%%%%%%%


\begin{thebibliography}{99}
\bibitem{footnote0} We illustrate this statement as follows. The reason why we can use $df = \frac{\partial f}{\partial z}dz + \frac{\partial f}{\partial z^*}dz^*$ is not because $dz$ and $dz^*$ are {\it linearly independent} differentials, but because $dz$ and $dz^*$ are {\it differentials}, that is {\it infinitesimal} quantities; they depend on each other 
	%$dz^*$ is in fact 100\% dependent on $dz$ since 
	simply because $dz^*$ is the complex conjugate of $dz$. Therefore we cannot argue that because $dz$ and $dz^*$ are independent variations, the partial derivatives $\frac{\partial f}{\partial z}$ and $\frac{\partial f}{\partial z^*}$ must be zero for the stationarity of $f$. There is a rather widespread misunderstanding of this mathematical fact to be found in the literature.
\bibitem{footnote1} When the variables and constraints are given in complex form, the method of Lagrange multipliers is expressed by 	$\frac{\partial f}
	{\partial z_k^*}
	=
	\sum_l
	\bigg[
	\lambda_l
	\frac{\partial g_l}
	{\partial z_k^*}
	+
	\lambda_l^*
	\frac{\partial g_l^*}
	{\partial z_k^*}
	\bigg]$. Though $g_l (z_k, z_k^*)=c$ and $g_l^* (z_k, z_k^*)=c^*$ are the same constraint, we have to add the complex conjugate of it in the Lagrange equation. Be careful to note that $(\frac{\partial g_l}{\partial z_k^*})^*$ is not equal to $\frac{\partial g_l^*}{\partial z_k^*}$. The above equation gives the condition for the stationarity of the real-valued function $f$.
\bibitem{footnote2} $\vec{m}_k^l$ indicates a short-hand notation of $\hat{a}_{l}^{\dag} \hat{a}_{k} | \vec{m} \rangle = \sqrt{(m_{l}+1)m_{k}} | \vec{m}_k^l \rangle$; that is, from the given configuration $\vec{m}$, one particle is removed from the $k$ orbital and one particle is added to the $l$ orbital, with the state $| \vec{m}_k^l \rangle$ being properly normalized.
\end{thebibliography}
\end{document}